\documentstyle[12pt]{article}
\font\mybb=msbm10 at 10pt
\def\bb#1{\hbox{\mybb#1}}

\begin{document}


\begin{titlepage}
\rightline{UB-ECM-PF-03/31}
\rightline{hep-th/0311159}

\vfill

\begin{center}
\baselineskip=16pt
{\Large\bf Fuzzy $C\!P^{(n|m)}$ as a quantum superspace{$^\star$}}
\vskip 0.3cm
{\large {\sl }}
\vskip 10.mm
{\bf ~Evgeny Ivanov$^{*,1}$, Luca Mezincescu$^{\dagger,2}$ and  Paul K.
Townsend$^{+,3}$}
\vskip 1cm
{\small
$^*$
Bogoliubov Laboratory of Theoretical Physics \\
JINR, 141980 Dubna, Russia\\
}
\vspace{6pt}
{\small
$^\dagger$
Department of Physics,\\
University of
Miami,\\
Coral Gables, FL 33124, USA\\
}
\vspace{6pt}
{\small

$^+$
Instituci\'o Catalana de Recerca i Estudis Avan\c cats,
\\
Departament ECM, Facultat de F\'{\i}sica, \\
Universitat de Barcelona,
Diagonal 647,\\
E-08028 Barcelona,
Spain
}
\end{center}
\vfill

\par
\begin{center}
{\bf
ABSTRACT}
\end{center}
\begin{quote}

The `Chern-Simons Quantum Mechanics'
of a particle on $C\!P^{(n|m)}$ is
shown to yield the fuzzy descriptions of
these superspaces, for which we
construct the non-(anti)commuting position
operators. For a particle on
the supersphere $C\!P^{(1|1)}\cong
SU(2|1)/U(1|1)$, the particle's
wave-function at fuzziness level $2s$ is
shown to be a degenerate
irrep of $SU(2|1)$ describing a supermultiplet of
$SU(2)$ spins
$(s-{1\over2},s)$.

\vfill
\vfill
\vfill

$^\star$ {To
appear in {\it
Symmetries in Gravity and Field Theory}, proceedings of
Conferencia
Homenaje en el 60 cumplea\~nos de Jos\'e Adolfo de Azc\'arraga,
June
9-11, 2003, Salamanca, Spain.}

\vfill
 \hrule width 5.cm
\vskip
2.mm
{\small
\noindent $^1$ eivanov@thsun1.jinr.ru\\
\noindent $^2$
mezincescu@server.physics.miami.edu\\
\noindent $^3$ pkt@ecm.ub.es. On leave
from Department of Applied
Mathematics and Theoretical Physics, University
of Cambridge, UK.
\\
}
\end{quote}
\end{titlepage}
\setcounter{equation}{0}
\section{Introduction}

Symplectic geometry provides a
general framework for classical mechanics
in which classical observables are
functions on a symplectic manifold $M$
(phase space) and their time
dependence is determined by the choice of
one of these observables as the
Hamiltonian. In the quantum theory, the
choice of an observable as the
Hamiltonian is again what determines
the time evolution of all other
observables, but the nature of these
observables is again independent of the
choice of Hamiltonian. If one
wishes to focus on the nature of the
correspondence between quantum and
classical observables it is therefore
possible, and convenient, to ignore
the Hamiltonian. In this case the
classical action is determined by the
symplectic 2-form $F$ on $M$.  As this
form is closed we can write it
locally as $F=dA$ for some abelian gauge
potential $A$ on $M$. The action
is then
\begin{equation}\label{action}
I =
s\int_w\!A
\end{equation}
where the integral is over some $1$-cycle $w$ (the worldline) in $M$, and
$s$ is a `coupling constant'. It would be possible to absorb $s$ into the
definition of $A$ but any multiple of $F$ determines a volume form
on $M$ and hence, for compact $M$, a total volume. We normalize $F$
by choosing $F/(4\pi\hbar)$ to give $M$ a unit total volume. The `actual'
volume of phase space is thus proportional to an appropriate power of
$s$. Note that although $M$ has a definite volume, it has no specified
{\sl metric}. The above action  defines a `topological' sigma-model with
`target space' $M$ that has been called `Chern-Simons-Quantum-Mechanics'
(CSQM) \cite{DJT,HT}. We shall follow this usage.

The symmetry group $G$ of the CSQM action
(\ref{action}) is the
infinite-dimensional group of symplectic
diffeomorphisms of $M$. To every
basis vector in the Lie algebra ${\cal G}$
of $G$ there is an associated
`moment map' from $M$ to ${\cal G}^*$, the
vector space dual to ${\cal
G}$. These are the Noether charges, and the
Poisson bracket turns the
vector space they span into a Lie algebra, which
is either homomorphic
to ${\cal G}$ or to a central extension of it. For
present purposes, the
important fact about the space ${\cal G}^*$ is that it
carries an
irreducible representation $R$ of $G$. This representation
decomposes into
an infinite sum of irreps of the maximal finite-dimensional
subgroup
$G_0$ of $G$. Consider, for example, $M=S^2$. In this case
$G_0=SO(3)$,
and the representation of $SO(3)$ carried by ${\cal G}^*$
is
\begin{equation}\label{sh}
R = {\bf 3}\oplus {\bf 5} \oplus {\bf 7}
\oplus \cdots \,.
\end{equation}
This is the $SO(3)$ representation content
in the expansion of a
non-trivial function on the 2-sphere in spherical
harmonics. The subspace
of ${\cal G}^*$ spanned by the triplet of Noether
charges is the
co-algebra of $G_0$ on which $G_0$ acts via its co-adjoint
action. The
co-adjoint orbits of $G_0$ in this space are 2-spheres
homothetic to
$M$.

In the quantum theory the Noether charges become
Hermitian operators
acting on a Hilbert space. They again span a Lie
algebra, which we will
call ${\cal G}_s^*$, but the Lie bracket is now $-i$
times the
commutator. This Lie algebra must be finite-dimensional if $M$
is
compact because a compact phase space implies a finite-dimensional
Hilbert
space and hence a finite number of linearly-independent
observables.
For present purposes, the important fact about ${\cal G}_s^*$
is that it
carries a representation $R_s$ of $G_0$ that is a truncation
of
the infinite sum of irreps carried by ${\cal G}^*$. For example,
if
$M=S^2$ then one can show that $2s$ must be an integer and the
Hilbert
space is a $(2s+1)$-dimensional carrier space for the spin $s$ irrep
of
$SU(2)$ \cite{berezin}. Note that the dimension of the Hilbert space
grows
linearly with $s$ as one would expect from the fact that $\hbar$
appears
in $I/\hbar$ only through the combination $s/\hbar$. The non
trivial
quantum observables in this case are the Hermitian $(2s+1)\times
(2s+1)$
matrices, excepting the identity matrix. They span a vector
space of
dimension $2s(2s+1)$ on which $SU(2)$, and hence
$SO(3)\cong
SU(2)/\bb{Z}_2$, acts reducibly according to the
representation
\begin{equation}\label{s2R}
R_s = {\bf 3} \oplus \cdots
\oplus ({\bf 4s +1})\,.
\end{equation}
The subspace of ${\cal G}_s^*$ on
which the triplet of $SU(2)$ acts
irreducibly plays a special role because
the hermitian operators that span
it are position operators for a `fuzzy'
sphere at fuzziness level $2s$
\cite{madore}. These operators generate the
associative algebra ${\cal
A}_{(2s+1)}$ of complex $(2s+1)\times (2s+1)$
matrices, which may be
considered as a finite-dimensional, but
non-commutative, approximation to
the infinite-dimensional commutative
algebra (with respect to ordinary
multiplication) of functions on $S^2$.

If
one views the 2-sphere as the space $C\!P^1\cong SU(2)/U(1)$ then
the
generalization to $C\!P^n\cong SU(n+1)/U(n)$ suggests itself. These
are
examples of K\"ahler manifolds, which provide a convenient supply
of
symplectic manifolds because the K\"ahler 2-form is a symplectic form.
In
this case, the gauge potential $A$ is determined, up to a
gauge
transformation, in terms of a K\"ahler potential $K$ via the
formula
\begin{equation}\label{Kform}
A = -i dz^i\, {\partial K \over
\partial z^i} + c.c.
\end{equation}
where $z^i$ ($i=1,\dots,n$) are $n$
complex local coordinates. For $C\!P^n$
we may
choose
\begin{equation}\label{kpotn}
K= \log \left(1 + \bar z\cdot
z\right),
\end{equation}
and the corresponding K\"ahler 2-form
is
\begin{equation}\label{K2}
F = -{2i d\bar z_i\wedge dz^i  \over (1+ \bar
z\cdot z) } + {2i(d \bar
z \cdot z)\wedge (\bar z\cdot d z) \over (1+ \bar
z\cdot z)^2 }\,.
\end{equation}
This 2-form has a manifest, linearly
realized, $U(n)$ invariance, but it
is also invariant under the non-linear
(and analytic)
infinitesimal
`translations'
\begin{equation}\label{analtran}
\delta z^i =
a^i + \left(\bar a \cdot z\right) z^i
\end{equation}
for $n$ complex
parameters $a^i$, as follows from the fact that
\begin{equation}
\delta A =
d\left(\bar a\cdot z + a\cdot \bar z\right).
\end{equation}
These
transformations close to yield a realization of the
Lie algebra $su(n+1)$
that exponentiates to a transitive action of
$SU(n+1)$ on $C\!P^n$.

The
$SU(n+1)$ invariance of $F$ implies the existence, via Noether's
theorem, of
$n(n+2)$ functions that span the Lie algebra $su(n+1)$ with
respect to the
Poisson bracket, which is
\begin{equation}\label{PB}
\{F,G\}_{PB} = {i\over
2s}\left(1+ \bar z\cdot z\right)
\left(\delta^i{}_j + z^i\bar z_j\right)
\left({\partial
F\over\partial z^i} {\partial G\over \partial \bar z_j} -
{\partial
F\over\partial \bar z_j} {\partial G\over \partial
z^i}\right)
\end{equation}
for any two functions $F,G$ on $C\!P^n$. The
vector space ${\cal G}^*$ of all
non-trivial functions on $C\!P^n$ is an
infinite-dimensional Lie algebra
with respect to this bracket, with
$su(n+1)$ as its maximal
finite-dimensional subalgebra. Its $SU(n+1)$
representation content
is
\begin{equation}
R = {\bf n(n+2)} \oplus {\bf
{1\over4}n(n+4)(n+1)^2} \oplus \cdots\,.
\end{equation}
This is a
generalization of (\ref{sh}). In the quantum theory this series
is truncated
to
\begin{equation}
R_s = {\bf n(n+2)} \oplus {\bf {1\over4}n(n+4)(n+1)^2}
\oplus \cdots
\oplus {\bf \frac{[(n+2s
-1)!]^2\,(n+4s)}{(n-1)!\,n!\,(2s!)^2}}\,,
\end{equation}
which generalizes
(\ref{s2R}). The $n(n+2)$ hermitian matrices form
an irrep of $SU(n+1)$ of
dimension $(n+2s)!/n!(2s)!$. These generate an
algebra of complex matrices
that defines the fuzzy versions of
$C\!P^n$
\cite{GS,TV,ABIY,BDL,AMMZ}.

Just as a fuzzy symplectic manifold
can be obtained from a CSQM model
model with a smooth symplectic manifold as
its target space, so
a fuzzy orthosymplectic {\it supermanifold} can be
obtained from a CSQM
model with an orthosymplectic target superspace. Again,
K\"ahler
supermanifolds provide a convenient supply of orthosymplectic ones,
the
simplest being the superspaces $\bb{C}^{(n|m)}$, for which the real
and
imaginary parts of the $m$ complex Grassmann-odd coordinates
become elements
of a Clifford algebra in the quantum theory. This
idea can be traced back to
the 1959 papers of Martin on the
mechanics of Grassmann variables
\cite{martin}. It also underlies the 1981
`quantum superspace' proposal of
Brink and Schwarz \cite{BS}. A
systematic study of non-(anti)commutative
deformations of Minkowski
superspace has been undertaken in recent years
(e.g. \cite{KPT,FLM})
and recent work on a string theory realization
(e.g.
\cite{dBGvN,OV,seiberg}) has initiated a revival of interest
in
non-(anti)commutative superspaces. 

In this paper we consider the class of
coset superspaces
\begin{equation}\label{scoset}
C\!P^{(n|m)} \cong
SU(n+1|m)/U(n|m)
\end{equation}
which all have $C\!P^n$ `body' and are both
homogeneous and
symmetric. They are also K\"ahler; in complex coordinates
\begin{equation}
Z^M = (z^i,\xi^\alpha) \qquad (i=1,\dots,n; \alpha= 1,\dots,m),
\end{equation}
with complex conjugates $\bar Z_M = (\bar z_i,\bar\xi_\alpha)$, the K\"ahler potential is
\begin{equation}\label{genpot}
K= \log \left(1+ \bar Z_M
Z^M\right)
\end{equation}
Note that $K$ is manifestly $U(n|m)$ invariant. The corresponding K\"ahler metric 
$ds^2 = dZ^Md\bar Z_N \partial^N\partial_M K$ (which, however,  is not used by the CSQM model) is invariant under the larger supergroup $SU(n+1|m)$, which is the maximal finite-dimensional sub-supergroup of the invariance supergroup of the  K\"ahler 2-form.  The principal new results of this paper concern the definition and study of the fuzzy, non(anti)commutative, versions of these superspaces. 
As we do this using CSQM methods, fuzzy $C\!P^{(n|m)}$ emerges as a `quantum superspace' in the sense of Brink and Schwarz.  One of our results is a construction of the
non-(anti)commuting position operators that generate the super-matrix algebra that defines the
fuzzy supergeometry of $C\!P^{(n|m)}$. As we shall see, there is no real distinction between
position and momentum operators in CSQM, but the position operator interpretation is natural in the semi-classical, large $s$, limit.

By an appropriate rescaling, in which one focuses only on
the
local properties of the target superspace, one can recover
$\bb{C}^{(n|m)}$
from $C\!P^{(n|m)}$, as we show in the final section of
this
paper. Here we
should raise a point that might have ocurred to the alert
reader. Darboux's
theorem states that any two symplectic manifolds of the
same dimension are
locally diffeomorphic, and the same applies to
orthosymplectic superspaces
as we shall illustrate in the final
section. Thus, if we are concerned only
with local properties then a
coordinate transformation suffices to take us
from the CSQM model with
target space $C\!P^{(n|m)}$ to one with target
space $\bb{C}^{(n|m)}$; there
is no
need to take a limit! The natural
K\"ahler metric on $C\!P^{(n|m)}$ is,
of course, quite different from the
flat K\"ahler metric on
$\bb{C}^{(n|m)}$,
but this is just a reflection of
the topological nature of the CSQM
action: it does not require a choice of
metric on the target space.
Under these circumstances one might wonder why
we do not begin by choosing
local coordinates for which the symplectic
2-form, and hence the CSQM
action, takes the simplest form. We will answer
this point in more detail
later, but the main reason is that the global
properties of $C\!P^{(n|m)}$
are most simply taken into account by viewing
it as a homogeneous coset
superspace for the supergroup $SU(n+1|m)$. An
additional advantage
is that the $SU(n+1|m)$ transformations are then
analytic.
This feature is actually essential to
the applicability of our
method, which was applied to the $n=0$ case
in \cite{IMPT}. The method has a
history of  relevance to this
conference, and we take the opportunity to
comment on it. We then
illustrate it with the $C\!P^n$ case before turning
to
$C\!P^{(n|m)}$.

The topological nature of the CSQM model with target
space $C\!P^{(n|m)}$
means that there is no dynamics; one can
consider the model `solved' when the nature of the Hilbert
space is
known. For the models with target superspace $C\!P^{(n|m)}$,
this
essentially amounts to a determination of the $SU(n+1|m)$
representation
content of the Hilbert space. This was worked out for the
special case
of $n=0$ in \cite{IMPT}. Here we consider the $n=1$ case; this
is of
particular interest because the $C\!P^{(1|m)}$ supermanifolds all
have
a $C\!P^1\cong S^2$ `body' and can therefore be viewed as
`$m$-extended
superspheres'.  The simplest of these, with real dimension
$(2|2)$ is
\begin{equation}
C\!P^{(1|1)}\cong
SU(2|1)/U(1|1)\,.
\end{equation}
We call this the `supersphere' because it
is the super-extension
of the sphere to a homogeneous symmetric K\"ahler
supermanifiold with
$SU(2)/U(1)\cong S^2$ `body' of minimal total
dimension\footnote{The term
`supersphere' has been used previously for the
coset superspace
$UOSp(1|2)/U(1)$ \cite{GKP,GR2,BKR,HIU,IU}. Although this
superspace is
often stated to have real dimension $(2|2)$, its `reality' is
defined with
respect to a `pseudoconjugation'; see e.g. \cite{AzA1} for
details. With
respect to standard complex conjugation, it actually has real
dimension
$(2|4)$ since spinors of $USp(2)\cong SU(2)$ span a vector space
of
dimension $4$ over the reals.}. We will show
that the `Hilbert' space of
a particle on a supersphere at fuzziness
level $2s$ is a degenerate irrep of
$SU(2|1)$ \cite{NRS} that decomposes
with respect to $SU(2)$ into a
supermultiplet of $SU(2)$ spins
$(s-{1\over2},s)$.

\setcounter{equation}{0}
\section{Analytic quantization of K\"ahler
models}

In local complex coordinates $z^i$ for K\"ahler target space $M$,
the
Lagrangian of the CSQM model with action (\ref{action})
is
\begin{equation}\label{lagone}
L= s\dot z^i A_i +
c.c.
\end{equation}
where $A_i$ are the complex components of the one-form
$A$, given in
terms of the K\"ahler potential $K$ by (\ref{Kform}). One
could pass to
the quantum theory by the usual Poisson bracket to
commutator
prescription, as outlined in the introduction, but here we
wish
to promote another method with some advantages that will
hopefully
become apparent as we proceed. This method
takes as its starting point the
alternative Lagrangian
\begin{equation}\label{equiv}
L = \left[p_i \dot z^i
+ \ell^i \varphi_i \right] + c.c.
\end{equation}
where $p_i$ are complex
momentum variables canonically conjugate to the
position space variables
$z^i$, and $\ell^i$ are complex Lagrange
multipliers that impose the complex
constraints\footnote{The
symbol $\approx$ indicates `weak' equality in
Dirac's sense.}
$\varphi_i\approx 0$,
with
\begin{equation}\label{constr}
\varphi_i = p_i - s A_i\,
.
\end{equation}
Solving the constraints for $p_i$ returns us to the
original
Lagrangian (\ref{lagone}), with phase space $M$, but the phase
space
of our alternative Lagrangian is, nominally, the cotangent bundle
of
$M$, so Poisson brackets of position variables now vanish.
However, the
constraints are `second-class', in Dirac's terminology,
and the standard way
of dealing with this is to replace Poisson
brackets by Dirac brackets. As
these are, by construction, equivalent
to the original Poisson brackets
derived from (\ref{lagone}), it might
seem that nothing has been
gained.

However, there is an alternative way of dealing with
second-class
constraints in these models. It arises from the observation
that the
complex functions $\varphi_i$ are in involution, as are their
complex
conjugates; it is only when we consider the two sets together that
the
constraints become second class. This suggests the possibility
of
regarding the constraints
$\varphi_i\approx 0$ as gauge fixing conditions
for gauge invariances
generated by the functions $\bar\varphi^i$. If we now
step back to the
un-gauge-fixed theory then Dirac tells us that we should
first quantize
{\it without constraint} by setting
\begin{equation}
p_i = -i
{\partial\over \partial z^i}\, , \qquad
\bar p^i = - i{\partial\over
\partial {\bar z}_i}\, ,
\end{equation}
which means that that the classical
constraint functions become the
quantum operators
\begin{equation}
\varphi_i
= -i\left[ {\partial\over \partial z^i} -s{\partial K\over
\partial z^i}
\right], \qquad
\bar\varphi^i = -i\left[{\partial\over \partial \bar z_i} +
s{\partial K
\over \partial \bar z_i} \right].
\end{equation}
We now ignore
the $\varphi_i$ constraints and treat their complex
conjugates as first
class by imposing the physical-state
conditions
\begin{equation}\label{physcon}
\bar\varphi^i|\Psi\rangle =0\,,
\qquad (i=1,\dots ,n)\,.
\end{equation}
This restricts physical
wavefunctions to take the form
\begin{equation}\label{physwave}
\Psi(z,\bar
z) = e^{-s K(z,\bar z)} \Phi(z)
\end{equation}
for {\it holomorphic}
function $\Phi$ of the $n$ complex coordinates of
$M$, which we shall call
the `reduced' wavefunction. As the K\"ahler
potential $K$ is not globally
defined (in general), this result shows
that $\Psi$ is not a true function\footnote{It is instead, see e.g.
\cite{Kirillov}, a section of a line bundle over $M$ with
curvature 2-form $sF/(2\pi)$, which must be integral for consistency.
Given our normalization of $F$, this requires requires $2s$ to be
an integer.}

A Hilbert space norm for $\Psi$ will take the
form
\begin{equation}
||\Psi||^2 = \int\! d\mu\, |\Psi|^2 = \int\! d\mu\
e^{-2sK}\, |\Phi(z)|^2
\end{equation}
where the integral is over $M$ and $d\mu$
is a measure on $M$. A natural
measure is provided by the volume form
determined by the appropriate
exterior power of the K\"ahler form $F$. In
particular, this defines a
$G_0$-invariant norm. Given this norm, one must
then further restrict the
physical states to be normalizable, and this
condition ensures that the
physical Hilbert space is
finite-dimensional.

For fermionic constraints, this alternative method of
dealing with
second-class constraints can be traced back to the 1976 papers
of
Casalbuoni \cite{casal} and papers in the early 1980s of Azc\'arraga et
al.
\cite{cowork,cowork2} and Lusanna \cite{LL}. A clear statement of it can
be
found, again for fermionic constraints, in a 1986 paper of
de
Azc{\'a}rraga and Lukierski \cite{AL}, who called it
`Gupta-Bleuler'
quantization by analogy with the procedure of that name for
covariant
quantization of electrodynamics\footnote{Noting that the Lorentz
gauge
condition cannot be consistently imposed as a physical state
condition,
Gupta and Bleuler suggested that it be separated into its
positive and
negative frequency parts (of which $\varphi$ and $\bar\varphi$
are
analogs) and that the positive frequency part be imposed as the
physical
state condition.}. It was also called Gupta-Bleuler quantization
in
the 1991 book of Balachandran et al. \cite{Bbook}, where it is
explained
for particle mechanics models with bosonic constraints. The
justification
for this method that we have sketched above arose in
independent work on
general models with bosonic second-class constraints
that can be separated
into two sets of {\it real} constraints, each in
involution \cite{MR,HM}.
In this context the method has become known as the
method of
`gauge-unfixing'. When, in a recent paper with Pashnev
\cite{IMPT}, we
advocated the holomorphic/anti-holomorphic variant of this
gauge-unfixing
procedure we were initially unaware of its earlier use and
chose to call
it the method of `analytic quantization' because the physical
state
conditions are essentially analyticity conditions.

Let us illustrate
the method for $M=C\!P^n$. Given the K\"ahler potential
of (\ref{kpotn}), we
deduce that physical wavefunctions have the form
\begin{equation}
\Psi = (1+
\bar z \cdot z)^{-s}\Phi(z)
\end{equation}
for holomorphic reduced
wavefunction $\Phi$. The K\"ahler 2-form
(\ref{K2}) determines the
$SU(n+1)$-invariant measure
\begin{equation}
d\mu = \prod_{i=1}^n (dz^i
d\bar z_i)\ (1+ \bar z \cdot z)^{-(n+1)} ,
\end{equation}
so the Hilbert
space norm is
\begin{equation}\label{Pnorm}
||\Psi||^2 = \prod_{i=1}^n
\int dz^i d\bar z_i\ (1+ \bar z \cdot
z)^{-(n +1 +2s)}\, 
|\Phi(z)|^2
\end{equation}
where the integral is over all values of $z$.
Normalizability of $\Psi$ thus requires $\Phi$
to be a
polynomial of maximum degree $2s$, from which it follows that $2s$
must
be a positive integer. The $(n+2s)!/[n!(2s)!]$ coefficients of
$\Phi$
span the irrep of $SU(n+1)$ formed from the $(2s)$-fold symmetric
tensor
product of the fundamental $({\bf n+1})$
irrep.

\setcounter{equation}{0}
\section{Position operators for fuzzy
$C\!P^n$}

Within the  `analytic', or `Gupta-Bleuler', method of
quantization, we
have a `large' Hilbert space in which position operators
commute, so that
the target space $M$ retains its status as a smooth,
classical,
configuration space. However, these `naive' position operators do
not act
on the physical subspace of Hilbert space because they take
physical
states into unphysical, or un-normalized, states. Thus, the
question
arises of how to construct the physical, non-commuting, position
space
operators. We will first address this issue for $M=C\!P^n$. In doing
so it
is useful to first consider how $C\!P^n$ can be embedded in a
Euclidean
space.

Any group acts naturally on the dual of its Lie algebra,
and the orbits of
this co-adjoint action are symplectic manifolds. Let $t_A$
span the Lie
algebra $su(n+1)$, such that
\begin{equation}
t_A t_B = {1\over
n+1}\, \delta_{AB} + {1\over \sqrt{2}} \left(d_{AB}{}^C
+ if_{AB}{}^C
\right) t_C
\end{equation}
where $f_{AB}{}^C$ are the structure constants;
$d_{ABC}\equiv
d_{AB}{}^D\delta_{CD}$ is the totally symmetric third-rank
invariant
tensor of $SU(n+1)$. The Kronecker delta $\delta_{AB}$ is a
Euclidean
metric, in cartesian coordinates $X^A$, on the
$n(n+2)$-dimensional vector
space dual to $su(n+1)$. The submanifold of this
space defined by
the constraints
\cite{BDL}
\begin{equation}\label{e-embed}
X^BX^A\delta_{AB}= {n\over n+1},
\qquad X^BX^A d_{AB}{}^C =
{\sqrt{2}(n-1)\over (n+1)} \,
X^C
\end{equation}
is a co-adjoint orbit of $SU(n+1)$ that is isomorphic to
$C\!P^n$.
Functions on $C\!P^n$ are therefore functions of the cartesian
coordinates
$X^A$ subject to these constraints. Thus, any (scalar) function
$\phi$ has
an expansion that starts as
\begin{equation}
\phi = \phi_0 +
\phi_A X^A + \phi_{AB}X^AX^B +\dots\, .
\end{equation}
The second term is
the ${\bf n(n+2)}$ (adjoint) irrep of $SU(n+1)$, so
these functions span its
Lie algebra with respect to the Poisson bracket
(\ref{PB}). The third term
might appear to contain all irreps appearing in
the symmetric product of two
adjoints, but only the
${\bf {1\over4}n(n+4)(n+1)^2}$ irrep survives the
constraints
(\ref{e-embed}).

We now turn to the quantum theory. The
`naive' position operators are just
multiplication by the complex
coordinates $z^i$ and $\bar z_i$. However,
these operators do not act on the
physical Hilbert space. Most obviously,
multiplication by $\bar z_i$ is not
physical because this operation fails
to commute with $\bar\varphi^i$.
Multiplication by $z^i$ {\it does} commute
with $\bar\varphi^i$, and takes a
holomorphic reduced wavefunction $\Phi$
to the new holomorphic reduced
wavefunction $z^i\Phi$, but if $\Phi$ is a
polynomial of maximal degree then
$z^i\Phi$ will not be normalizable.
Thus, multiplcation by $z^i$ is not
physical either. We therefore need to
modify the `naive' position operators.
As $\hbar$ occurs only through the
combination $\hbar/s$, and we choose
units such that $\hbar=1$, the
dimensionless number $1/s$ effectively plays
the role of $\hbar$. We thus
expect quantum corrections to be of order
$1/s$. As the classical
coordinates commute, we thus seek physical  position
operators $\hat z^i$
of the form
\begin{equation}\label{semic}
\hat z^i =
z^i + {\cal O}(1/s).
\end{equation}
The quantum correction must be such that
these operators, and their
hermitian conjugates (defined via the Hilbert
space norm) commute weakly
with the operators $\bar\varphi^i$ that
annihilate physical states. This
condition has the unique
solution:
\begin{equation}
\hat z^i = z^i - {1\over s} \left[ {\partial
\over \partial
\bar z_i} +  z^i \left(z \cdot {\partial \over
\partial
z}\right)\right], \qquad
\hat{\bar z}_i = \bar z_i + {1\over s}
\left[ {\partial \over \partial
z^i} + \bar z_i \left(\bar z \cdot {\partial
\over \partial \bar
z}\right)\right].
\end{equation}
The effect of the
physical position operators on the reduced
wavefunction
is
\begin{equation}
\hat z^i : \Phi \rightarrow s^{-1}u^i \Phi,
\qquad
\hat{\bar z}_i: \Phi \rightarrow s^{-1}\bar u_i
\Phi
\end{equation}
where
\begin{equation}\label{redops}
u^i = z^i\left(2s
- z\cdot {\partial \over \partial z}\right),
\qquad \bar u_i = {\partial
\over \partial z^i}\, ,
\end{equation}
from which it follows immediately
that a normalized reduced wavefunction
is taken to another normalized
reduced wavefunction. Acting on the
coordinates $z^i$, the operators $u,\bar
u$ generate the infinitesimal
non-linear translations (\ref{analtran}). In
other words, the
operators $s\hat z^i$ are {\it momentum} operators for a
particle on
$C\!P^n$. There is no real distinction between the position
and
momentum operators. For large $s$ the position operator interpretation
is the natural one, because of (\ref{semic}), whereas the momentum
operator
interpretation is natural for small $s$. Thus, CSQM models
exhibit a kind of
duality between the semi-classical and ultra-quantum
regimes in which
position and momentum are interchanged.

Note that $\hat z^i$ and $\hat
z^j$ commute, but
\begin{equation}\label{commz}
[\hat z^i,\hat{\bar z}_j] =
{1\over s^2}\left[ {\hat J}^i{}_j +
\left({n+1\over n}\right)\delta^i{}_j
\hat J_0\right]
\end{equation}
where
\begin{equation}
{\hat J}^i{}_j =
z^i{\partial\over \partial z^j} - \bar z_j{\partial\over
\partial \bar z_i}
- {1\over n} \left( z\cdot {\partial\over \partial z} -
\bar z\cdot
{\partial\over \partial \bar z}\right)\delta^i_j
\end{equation}
is the
generator of $SU(n)$, and
\begin{equation}\label{uone}
\hat J_0 = z\cdot
{\partial\over \partial
z} - \bar z\cdot {\partial\over \partial \bar z} -
2s\, {n\over n+1}
\end{equation}
is the $U(1)$ generator. As expected, the
physical quantum position
operators are non-commutative, but the
non-commutativity disappears in the
classical limit $s\rightarrow \infty$.
The shift in the $U(1)$ charge by a
term proportional to $s$ is due to the
fact that the Lagrangian is not
invariant under (\ref{analtran}) but changes
by a total derivative. This implies that the Hilbert space will carry a {\it projective} representation of $SU(n+1)$ rather than a true representation\footnote{We pass over the associated subtleties here, but note that this has been a major theme of Jose-Adolfo de Azc\'arraga's work; we refer the reader to his bool with J-M Izquierdo \cite{AI}.} 

The commutation relation (\ref{commz}) has an analog
for the operators
$u,\bar u$, acting on reduced wavefunctions.
Specifically,
\begin{equation}\label{comm1}
[u^ii,\bar u_j] = {\hat j}^i{}_j
+ \left({n+1\over n}\right)\delta^i{}_j
{\hat
j}_0
\end{equation}
where
\begin{equation}
{\hat j}^i{}_j =
z^i{\partial\over\partial z^j} - {1\over n}
\delta^i{}_j \, \left(z\cdot
{\partial\over\partial z} \right) \, ,\qquad
{\hat j}_0 = z\cdot {\partial \over
\partial z} - 2s\, {n\over n+1}
\end{equation}
are the $SU(n)$ and $U(1)$
operators acting on reduced wavefunctions.
The commutation relations of
these operators with $u,\bar u$ imply that
the latter have $U(1)$ charges
$1,-1$, and transform as the
${\bf n},\bar{\bf n}$ of $SU(n)$, respectively.
The same conclusion
results, with more effort, from the commutation
relations of $\hat
J^i{}_j$ and $\hat J_0$ with $\hat z^i$ and $\hat{\bar
z}_i$. Thus,
whether they act on the `full' Hilbert space or the reduced
Hilbert
space, {\it the physical position operators of a fuzzy $C\!P^n$
are
proportional to the ladder operators of a representation of
$SU(n+1)$.}

This result has a classical counterpart that is related to our
earlier
description of $C\!P^n$ as a co-adjoint orbit of $SU(n+1)$. The
functions
on phase space corresponding to the physical position operators
$\hat z^i$
are\footnote{This defines a transformation of the
coordinates within the `large' phase space. See \cite{LM} for
a discussion of this point in non-commutative quantum mechanics,
motivated by results in \cite{CFZ,DBLS}.}
\begin{equation}
z^i -{i\over s} \left[\bar
p^i + z^i \left(z\cdot
p\right)\right] = {2z^i \over 1 + \bar z\cdot z}
-{i\over
s}\left[\bar\varphi^i + z^i\left(
z\cdot\varphi\right)\right]
\approx
{2z^i \over 1 + \bar z\cdot
z}\,.
\end{equation}
The functions corresponding to the rescaled quantum
position
operators $s \hat z^i$ are therefore
\begin{equation}
w^i =
2s\left(1 + \bar z\cdot z\right)^{-1} z^i\,.
\end{equation}
The Poisson
bracket (\ref{PB}) of $w^i$ with $w^j$ is zero
but
\begin{equation}
i\{w^i,\bar w_j\}_{PB} =  J^i{}_j + \, \left(n+1\over
n\right)\,
\delta^i{}_j
J_0
\end{equation}
where
\begin{equation}
J^i{}_j =
{2s\over 1+ \bar z\cdot z}\left(z^i\bar z_j - {1\over
n}\delta^i{}_j
\bar
z\cdot z\right) , \qquad
J_0 = {2s\over n+1}\left({\bar z\cdot z - n \over 1
+ \bar z\cdot
z}\right).
\end{equation}
These are the classical generators
of $U(n)$, with Poisson brackets
\begin{equation}
i\{J^i{}_j,J^k{}_l\}_{PB}
=  \left(J^i{}_l \delta^k{}_j - J^k{}_j
\delta^i{}_l\right), \qquad
\{J^i{}_j,J_0\}_{PB} =0\,.
\end{equation}
In
addition,
\begin{eqnarray}
i\{J^i{}_j,w^k\} = w^i \delta^k{}_j - {1\over
n}\delta^i{}_j w^k\, , &&
i\{J_0,w^i\} = w^i, \nonumber \\
i\{J^i{}_j,\bar
w_k\} = -\bar w_j \delta^i{}_k + {1\over n} \delta^i{}_j
\bar w_k, &&
i\{J_0,\bar w_i\} = - w_i\,.
\end{eqnarray}
The functions $(w,\bar w,J,J_0)$
thus span a Lie algebra, with the
Poisson bracket as the algebra product.
This is the Lie algebra of
$SU(n+1)$. It follows that $(w,\bar w,J,J_0)$ are
linear
combinations of the real embedding space coordinates $X^A$ that
we
introduced previously. The precise relation can be found by comparing
the
identity
\begin{equation}
w^i\bar w_i + {1\over2}J^i{}_j J^j{}_i +
{n+1\over 2n}\, J_0^2 \equiv
{2n\over n+1}\, s^2
\end{equation}
with the
first equation in (\ref{e-embed}).

\setcounter{equation}{0}
\section{Fuzzy
$C\!P^{(n|m)}$}

We now turn to the case of $C\!P^{(n|m)}$, for which the
K\"ahler potential
is given in (\ref{genpot}). The corresponding CSQM
Lagrangian is
\begin{equation}\label{CQSMlag}
L = -is \left(1+ \bar Z\cdot
Z\right)^{-1} \bar Z \cdot \dot Z +c.c.
\end{equation}
and the K\"ahler
2-form is
\begin{equation}\label{K22}
F = -{2i dZ^M\wedge d\bar Z_M \over
(1+ \bar Z\cdot Z)}  + {2i(d
\bar z^M \cdot Z)\wedge (\bar Z\cdot d
Z)\over
(1+ \bar Z\cdot Z)^2}\,.
\end{equation}
Apart from the manifest
$U(n|m)$ invariance, this 2-form is also invariant
under the infinitesimal,
and analytic, transformation
\begin{equation}\label{analytic}
\delta Z^M =
a^M + (\bar a \cdot Z) Z^M
\end{equation}
where $a^M= (a^i,\epsilon^\alpha)$
for commuting parameters $a^i$ and
anticommuting parameters
$\epsilon^\alpha$. These transformations
close to yield a realization of the
Lie superalgebra $su(n+1|m)$ that
exponentiates to the action of $SU(n+1|m)$
on $C\!P^{(n|m)}$. We can
therefore identify these spaces as the coset
superspaces
$SU(n+1|m)/U(n|m)$, as claimed in the introduction.

To make
this identification precise, we should clarify how the superalgebra
$u(n|m)$
is embedded in $su(n+1|m)$. Let $Q^a_\alpha$  ($a,b = 1,
\ldots
n+1;
\alpha,\beta = 1,\ldots m$) be the odd $su(n+1|m)$ charges.
Their
anticommutation relations are
\begin{equation}
\{ Q^a_\alpha, \bar
Q^\beta_b \} = \delta^a{}_b I^\beta{}_\alpha +
\delta^\beta{}_\alpha T^a{}_b
+ \delta^a{}_b \delta^\beta{}_\alpha\,
B\qquad (I^\alpha{}_\alpha = T^a{}_a
=0)
\label{1}
\end{equation}
where $I^\beta{}_\alpha$, $T^a{}_b$ and $B$
are
the generators of the mutually commuting $su(m)$, $su(n+1)$ and
$u(1)$
algebras, which form the even sub-algebra of $su(n+1|m)$.
The odd charges
have the following commutation
relations with the $su(m)$
generators:
\begin{equation}
[I^\alpha{}_\beta, Q^a_\gamma] =
\delta^\alpha{}_\gamma Q^a_\beta -
\frac{1}{m}\delta^\alpha{}_\beta
Q^a_\gamma\,, \quad
[I^\alpha{}_\beta, \bar Q^\gamma_a] =
-\delta^\gamma{}_\beta \bar
Q^\alpha_a + \frac{1}{m}\delta^\alpha{}_\beta
\bar Q^\gamma_a\, .
\end{equation}
Their commutation relations with the
$su(n+1)$ generators
$T^a{}_b$ are similar, but with $1/m \rightarrow
1/n+1$.
The Jacobi identities uniquely fix the commutation relations
with
the $u(1)$ generator to be
\begin{equation}
[B, Q^a_\beta] =
\left(\frac{1}{m} - \frac{1}{n+1}\right)Q^a_\beta\;,
\quad [B, \bar
Q^\beta_a] =
- \left(\frac{1}{m} - \frac{1}{n+1}\right)\bar
Q^\beta_a\;.
\end{equation}

Now let us see how the subalgebra $u(n|m)$ of
$su(n+1|m)$ is singled out.
Let
\begin{equation}
Q^a_\beta = (Q^i_\beta,
S_\beta), \quad \bar Q^\beta_a =
(\bar Q^\beta_i, \bar S^\beta)\qquad (i =
1, \ldots n).
\end{equation}
Then the basic anticommutator of $su(n|m)$
is
\begin{equation}
\{ Q^i_\beta, \bar Q^\alpha_j \} =
\delta^i{}_j
I^\alpha{}_\beta + \delta^\alpha{}_\beta
\tilde{T}^i{}_j +
\delta^i{}_j
\delta^\alpha{}_\beta\, \tilde
B\,
\end{equation}
where
\begin{equation}
\tilde{T}^i{}_j = T^i{}_j
+
{1\over n}\delta^i{}_j J\,, \quad
\tilde B = \left(B -\frac{1}{n}J
\right), \quad J \equiv
T^{n+1}{}_{n+1}\,.
\end{equation}
The new `internal'
$U(1)$ charge $\tilde{B}\subset su(n|m)$ has
the following commutation
relations with the odd charges:
\begin{equation}
[\tilde B, Q^i_\alpha]
=
\left(\frac{1}{m} - \frac{1}{n}\right)Q^i_\alpha\,.
\end{equation}
We
could define $u(n|m)$ by adding the generator $B$ to those
of $su(n|m)$,
with which it forms a semi-direct sum, but this would
clearly not lead to a
coset superspace with body $SU(n+1)/U(n)\cong C\!P^n$.
Instead we add the
generator $J = T^{n+1}{}_{n+1}$, which has the
following comutation
relations with the odd charges:
\begin{equation}
[J, Q^i_\alpha] = -
\frac{1}{n+1}\,Q^i_\alpha\,.
\end{equation}
This $u(n|m)$ subgroup is again
a semi-direct sum of $u(1)$ with
$su(n|m)$, and it yields a coset superspace
with $C\!P^n$ body.

Superfields on $C\!P^{(n|m)}$ can be viewed as unitary
irreps of
$SU(n+1|m)$ induced from an unitary representation of its
$U(n|m)$
subgroup, but we may limit ourselves to representations of $U(n|m)$
that
are $SU(n|m)$ singlets with non-zero $U(1)$ charge $J$.
Since the
`matrix' part of $\tilde B$ is trivial for such $SU(n|m)$
singlets, the
natural $U(1)$ charge labelling the superfields is $J
\subset su(n+1)$ (its
`matrix' part); as we shall see, this charge is
just $2s$.

We now turn to
the quantum theory of the CSQM model with $C\!P^{(n|m)}$ as
target space. As
for the $C\!P^n$ case, we expect the quantum position
operators to take the
form
\begin{equation}
\hat Z^M = Z^M + {\cal O}(1/s)
\end{equation}
with the
quantum corrections such that both the operators $\hat Z^M$, and
their
Hermitian conjugates with respect to the norm (\ref{gennorm1}), act
on the
physical subspace of `Hilbert space'. There is again a unique
solution to
this problem, and one finds that
\begin{eqnarray}
\hat Z^M &=& Z^M - {1\over
s} \left[ {\partial \over \partial
\bar Z_M} +  Z^M \left(Z \cdot {\partial
\over \partial
Z}\right)\right] \approx s^{-1} W^M \nonumber \\
\hat{\bar
Z}_M &=& \bar Z_M + {1\over s} \left[ {\partial \over \partial
Z^M} + \bar
Z_M\left(\bar Z \cdot {\partial \over \partial \bar
Z}\right)\right] \approx
s^{-1} \bar W_M
\end{eqnarray}
where
\begin{equation}
W^M = 2s \left(1+ \bar
Z\cdot Z\right)^{-1} Z^M.
\end{equation}
The classical functions $W^M$ and
their complex conjugates have Poisson
brackets that close on the functions
generating the manifest $U(n|m)$
symmetry of the Lagrangian, and together
these span the Lie superalgebra
$su(n+1|m)$ with respect to the
Poisson
bracket determined by the othosymplectic 2-form $F$. In the
quantum
theory, the rescaled quantum position operators $s\hat Z^M$
and
$s\hat{\bar Z}_M$ act on the reduced wavefunction via the
operators
\begin{equation}
U^M = Z^M\left(2s - Z\cdot {\partial \over
\partial Z}\right),
\qquad \bar U_M = {\partial \over \partial Z^M}\,
,
\end{equation}
which generalize the corresponding operators (\ref{redops})
of the $C\!P^n$
case. These operators close on $su(n+1|m)$. Thus
{\it the
physical position operators of a fuzzy $C\!P^{(n|m)}$ are
proportional to
the ladder operators of a representation of $SU(n+1|m)$},
a result that
precisely parallels the one we found for $C\!P^n$.

We now turn to the
nature of the physical states, and their representation
content. Proceeding
exactly as in the $C\!P^n$ case, one finds that physical
wavefunctions take
the form
\begin{equation}
\Psi = \left[1 + \bar Z \cdot
Z\right]^{-s}\Phi(Z)
\end{equation}
for holomorphic superfield $\Phi(Z)$.
The `Hilbert space norm' is
\begin{eqnarray}\label{gennorm1}
||\Psi||^2 &=&
\int\! d\mu_0 \left[1 + \bar Z\cdot Z\right]^{m-n-1}
|\Psi|^2 \nonumber \\
&=&
\int\! d\mu_0 \left[1 + \bar Z \cdot Z\right]^{m-n-1-2s}
|\Phi|^2
\end{eqnarray}
where (allowing for an arbitrary normalization factor ${\cal
N}$)
\begin{equation}
d\mu_0 = {\cal N}\prod_{i=1}^n dz^i d\bar
z_i \prod_{\alpha=1}^m
{\partial\over
\partial \bar\xi_\alpha}
{\partial\over \partial \xi^\alpha}\, .
\end{equation}
If $n\ne 0$ then
normalizability of $\Psi$ requires $\Phi$ to be a
polynomial in $Z$ of
maximal degree $2s$. The case $n=0$ is
special because all coordinates are
Grassmann-odd. There is no
quantization condition on $2s$ in this case but
special features emerge
for particular integer values of $2s$. Generically,
the `Hilbert' space
transforms as a reducible but not fully reducible
representation of
$SU(1|m)$ but for certain integer values of $2s$ there are
zero norm
states, and this requires a redefinition of the physical
`Hilbert'
space as an equivalence class of states modulo the addition of
zero norm
states.  This redefined physical subspace is an $SU(1|m)$ singlet
when
$2s=m-1$. When $2s=m-2$ (assuming $m>2$) the redefined
physical
`Hilbert' space is $(m+1)$-dimensional and transforms as the
fundamental
irrep of $SU(1|m)$. We refer to \cite{IMPT} for
details\footnote{The
supergroup called $SU(1|m)$ here was called $SU(m|1)$
in \cite{IMPT}, and
we deviate from that reference in a few other minor
respects too. Note
that the parameter $\alpha$ of \cite{IMPT} can be chosen
to equal the
parameter $\gamma$ which should then be identified with the
parameter $2s$
of this paper.}.

We will pass over the issue of the
$SU(n+1|m)$ content of `Hilbert' space
for the general $n\ne0$ case, and
instead concentrate on $n=1$.
As mentioned in the introduction, this case
corresponds to a particle on
the $m$-extended supersphere. We begin with
$m=1$; a particle on the
(simple) supersphere. In this case, the `Hilbert'
space norm
(\ref{gennorm1}) is
\begin{eqnarray}\label{gennorm2}
||\Psi||^2 =
\int\! d\mu_0 \left[1 + \bar Z \cdot Z\right]^{-(1+2s)}\,
|\Phi|^2
\end{eqnarray}
where $\Phi(z,\xi)$ is the `reduced' holomorphic
wavefunction
corresponding to $\Psi$. It has the superfield
expansion
\begin{equation}
\Phi = \phi_0(z) + \xi
\phi_1(z)
\end{equation}
where $\phi_0$ and $\phi_1$ are two holomorphic
functions of opposite
Grassmann parity. After performing the Berezin
integrals over
$\xi$ and $\bar\xi$, one finds
that
\begin{equation}
||\Psi||^2 \, \propto \, \left[ \int_{S^2} {dz d\bar z \over
(1+ \bar z
z)^{2s+1}}\, |\phi_1|^2 - (1+2s) \int_{S^2} {dz d\bar z \over (1+
\bar z
z)^{2s+2}}\, |\phi_0|^2 \right].
\end{equation}
The relative minus
sign between the two integrals does not imply that the
norm is indefinite
because we can choose the overall normalization such
that the offending
integral is the one with the nilpotent integrand.
Leaving aside this issue,
normalizability of the second integral implies
that $\phi_0(z)$ is a
polynomial of maximum degree $2s$ and hence that its
coefficients transform
as spin $s$ under $SU(2)$, while normalizability of
the first integral
implies that $\phi_1(z)$ is a polynomial in $z$ of
maximum degree $(2s-1)$,
and hence that its coefficients transform as spin
$s-{1\over2}$ under
$SU(2)$. If we wish to respect the spin-statistics
connection\footnote{This
is not a mathematical necessity here because
the spins are
non-relativistic.} then we should choose $\Phi$ to have
Grassmann parity
$(-1)^{2s}$. We then have a `Hilbert' space that is a
supermultiplet with
spins $\left(s -{1\over 2}, s\right)$
carrying a ${\bf 2s}\oplus ({\bf
2s+1})$ representation
of $SU(2)$; this is the decomposition into $SU(2)$
irreps of the
`degenerate' irrep of $SU(2|1)$ of total dimension $4s+1$
\cite{NRS}.

When $m>1$ we have a particle on the $m$-extended supersphere.
An
analysis along the same lines leads to an $m$-extended supermultiplet
of
$SU(2)$ irreps with multiplicities given by binomial coefficients,
which is
reminiscent of particle supermultiplets of the super-Poincar\'e
group. For
example, when
$m=2$ and $2s=1$ we get the supermultiplet of spins
$\left(0,
{1\over2},{1\over2}, 1\right)$, which could be viewed as a
non-relativistic
analogue of the vector supermultiplet of $N=2$
super-Poincar\'e
supersymmetry. The supergroups $SU(2|m)$ can thus be viewed
as
non-relativistic supersymmetry groups, their degenerate
representations
yielding supermultiplets of $SU(2)$ spins. In the simplest,
$m=1$,
$2s=1$, case we have a supermultiplet of spins $(0,1/2)$, described
by a
Grassmann-even scalar wavefunction and a Grassmann-odd
Pauli-spinor
wavefunction
respectively.

\setcounter{equation}{0}
\section{The local limit}

Let us
define new coordinates $V^M$ for $C\!P^{(n|m)}$ by
setting
\begin{equation}
Z^M =  \left(1-\bar V\cdot V\right)^{-{1\over2}}\,
V^M\,.
\end{equation}
One may verify, by substitution,
that
\begin{equation}\label{darboux}
F= -2i\, d\bar V_M \wedge
dV^M\,,
\end{equation}
and hence that the Lagrangian in the new coordinates
is
\begin{equation}\label{simplag}
L = -is\, \bar V \cdot \dot V +
c.c.\,.
\end{equation}
This illustrates the extension to orthosymplectic
supergeometry of
Darboux's theorem in symplectic geometry, which states that
all
symplectic manifolds of the same dimension are {\it locally}
equivalent.
In this case the equivalence is between $C\!P^{(n|m)}$, with its
K\"ahler
2-form as the symplectic form, and $\bb{C}^{(n|m)}$ with its
orthosymplectic
2-form determined by the natural complex structure. The
superspace
$C\!P^{(n|m)}$ differs {\it globally} from $\bb{C}^{(n|m)}$ by
the
restriction to the `superball' in $\bb{C}^{(n|m)}$ with
the `unit
sphere' $\bar V\cdot V=1$ as its
boundary; note that this boundary
corresponds to infinity in the $Z$
coordinates.

The natural complex
structure on $C\!P^{(n|m)}$ differs from the one
implicit in the choice of
$Z$ coordinates on $C\!P^{(n|m)}$ because $Z$ is
not a holomorphic function
of $V$. This has the consequence that a
hermitian metric in the $Z$
coordinates will not be hermitian in the $V$
coordinates, and vice-versa.
Thus, the hermitian (K\"ahler) metric $g$
that one may naturally associate
with the K\"ahler 2-form $F$ via
the
formula
\begin{equation}\label{Fmetric}
F = -2i d\bar Z^M g_M{}^N \wedge
dZ_N
\end{equation}
is not diffeomorphic to the metric that this formula
yields in the
$V$ coordinates, because $g$ is not hermitian in these
coordinates. The
K\"ahler metric in the $V$ coordinates defined by the
analogous
formula is a {\it flat} metric on $\bb{C}^{(n|m)}$. This
illustrates the
topological nature of CSQM: the action does not involve a
choice of metric
on the target space $M$. However, the quantum theory
does
involve a choice of complex structure (in our approach this choice
is
implicit in the separation of the second class constraints
into
holomorphic and anti-holomorphic pairs) and once a complex structure
has
been chosen then it is preserved only by holomorphic diffeomorphisms.
The
holomorphic diffeomorphisms that leave $F$ invariant are
holomorphic
symplectic diffeomorphisms, generated by holomorphic functions
on $M$.

One reason that we chose to consider the CSQM model for target
space
$C\!P^{(n|m)}$ in the $Z$ coordinates is that the
transformations
(\ref{analytic}) are analytic in these coordinates. They are
not
analytic in the $V$ coordinates. Thus, while one can always put the
CSQM
action in the form (\ref{simplag}) by a change of coordinates, one
cannot
do this with a {\it holomorphic} change of coodinates. This makes the
extension of Darboux's theorem to the quantum theory
problematic; not surprisingly in view of the non-local features of
quantum mechanics\footnote{For example, in the path-integral approach
one must consider amplitudes for all paths in phase space, so the global
features of phase space can affect local physics.}. Nevertheless, one can
still focus on the local properties of the target space by introducing
rescaled coordinates
${\cal Z}^M$ such that
\begin{equation}
Z^M = {1 \over \sqrt{s}}\, {\cal Z}^M\,.
\end{equation}
In the limit that $s\rightarrow \infty$ we have
\begin{equation}
s\log \left(1+ \bar Z\cdot Z\right) \rightarrow  \bar {\cal Z}\cdot
{\cal Z}\,,
\end{equation}
which is the K\"ahler potential for $\bb{C}^{(n|m)}$. Physical wavefunctions
now take the coherent state form
\begin{equation}
\Psi = e^{-\bar {\cal Z}\cdot {\cal Z}}\, \Phi({\cal Z})\,.
\end{equation}
The rescaled physical position operators are now
\begin{equation}
\hat {\cal Z}^M = {\cal Z}^M - {\partial \over
\partial \bar {\cal Z}_M}\, ,\qquad
\hat {\bar {\cal Z}}_M = \bar {\cal Z}_M +  {\partial \over \partial
{\cal Z}^M},
\end{equation}
for which the non-zero (anti)commutation relations are
\begin{equation}
[\bar {\cal Z}_M, {\cal Z}^N\} = 2\delta_M{}^N\,.
\end{equation}
In other words, the physical position operators obey the (anti)commutation
relations of the non-(anti)commutative $\bb{C}^{(n|m)}$.

\bigskip
\noindent
{\bf Acknowledgements.} We thank Jos\'e-Adolfo de Azc\'arraga
and Joaquim Gomis for helpful discussions. E.I. thanks the Department
of Physics of the University of Padua for
warm hospitality during the final stages of this study. L.M. thanks the
theory group
at JINR for hospitality and partial financial support, and the organisers
of the {\sl Workshop on Branes and Generalized Dynamics} (Argonne, October
20-24, 2003) for the opportunity to present some of the results of this
paper. The work of E.I. was supported in part by grants DFG No.436 RUS
113/669,
RFBR-DFG 02-02-04002, INTAS 00-00254, RFBR 03-02-17440 and a grant of the
Heisenberg-Landau program. L.M. was supported in part by the National
Science
Foundation under grant PHY-9870101.

\end{document}